\documentclass[twocolumn]{IEEEtran}
\usepackage{amsmath,amssymb,graphicx,color,enumitem,hyperref}
\pdfoutput=1
\usepackage{fancyhdr}
\def\mathlette#1#2{{\mathchoice{\mbox{#1$\displaystyle #2$}}%
                               {\mbox{#1$\textstyle #2$}}%
                               {\mbox{#1$\scriptstyle #2$}}%
                               {\mbox{#1$\scriptscriptstyle #2$}}}}
\renewcommand{\vec}[1]{\mathlette{\boldmath}{#1}}
\newcommand{\prob}[1]{\text{Pr}\left[#1\right]}
\newcommand{\expect}[2][]{\mathbb{E}_{#1}\left[#2\right]}
\newcommand{\pchk}{\mathrm{H}}
\DeclareMathOperator\weight{wt}

\newcommand{\set}[1]{\mathcal{#1}}

\newcommand{\figwidth}{0.84\columnwidth}

\title{Estimating Channel Parameters from\\ the Syndrome of a Linear Code}
\author{Gottfried Lechner and Christoph Pacher\thanks{\hrule \vspace{2mm}
	This work has been funded by the Vienna Science and Technology Fund (WWTF) through project ICT10-067 (HiPANQ).
	G.~Lechner is with the Institute for Telecommunications Research at the University of South Australia (\href{mailto:gottfried.lechner@unisa.edu.au}{gottfried.lechner@unisa.edu.au}).
	C.~Pacher is with the Safety \& Security Department at the Austrian Institute of Technology (\href{mailto:christoph.pacher@ait.ac.at}{christoph.pacher@ait.ac.at}).
}
}

\graphicspath{{figures/}}

\begin{document}

\maketitle
\thispagestyle{fancy}

\begin{abstract}
In this letter, we analyse the properties of a maximum likelihood channel estimator based on the syndrome of a linear code. For the two examples of a binary symmetric channel and a binary input additive white Gaussian noise channel, we derive expressions for the bias and the mean squared error and compare them to the Cram\'er-Rao bound. The analytical expressions show the relationship between the estimator properties and the parameters of the linear code, i.e., the number of check nodes and the check node degree.
\end{abstract}


\section{Introduction}
Channel state information (CSI) at the receiver, i.e., knowledge of parameters like the crossover probability or the signal-to-noise ratio, is often assumed when discussing forward error correction. CSI can be of interest for various reasons:
\begin{itemize}[leftmargin=5mm]
\item When using decoding algorithms such as the sum-product algorithm, CSI is required to compute log-likelihood ratios. For most decoding algorithms, absence or inaccuracy of CSI results in a higher probability of a decoding error \cite{Saeedi2007}.
\item If the receiver has knowledge of the CSI then it can predict whether a decoding attempt will be successful (e.g., by comparing the channel parameter to a threshold). This information can be used to request additional data in an automatic repeat request (ARQ) system or to discard the received block and save energy by not even attempting to decode.
\end{itemize}

In this paper we analyse estimation of CSI based on the syndrome of a linear code. This is an interesting task as it does not involve additional calculations at the sender nor any communication overhead.

To compute the syndrome, the receiver performs a hard decision on the received signal, thereby converting the channel to a binary symmetric channel (BSC). The channel state information for the original channel is then derived from the estimated cross over probability of this BSC.

Of special importance is the case when the actual channel is a BSC. The reason for this is that the channel outputs of a BSC (with uniform input distribution) do not carry any information about the error rate of the BSC. Hence, there are no estimation techniques which can ignore the code constraints and operate directly on the channel outputs.

In addition to the BSC we present results for the binary input additive white Gaussian noise (BI-AWGN) channel. 

The idea of estimating channel parameters based on the syndrome of a linear code is not new. In \cite{TotoZarasoa:2011p13281}, the authors present estimation of the BSC parameter of a Slepian-Wolf problem (and the modification for standard channel coding). We extend these results by providing a simple analytic expression for the estimator, and expressions for bias, mean squared error (MSE) and the Cram\'er-Rao bound which enables us to study the influence of the system parameters in detail. Additionally, we extend the results to the BI-AWGN case.

In \cite{Chen:2012ki}, the authors present \emph{error estimating coding} that enables the receiver to estimate the bit error probability of received packets. Assuming codewords of length $n$, they prove that $\mathcal{O}\left(\log(n)\right)$ bits have to be added to obtain an estimate that exceeds a maximum relative error $\epsilon$ with at most probability $\delta$. Estimating the bit error probability based on \emph{sketch data structures} is presented in \cite{Hua:2012cx}.
The complexity of their scheme is lower than in \cite{Chen:2012ki} but the asymptotic behaviour is still $\mathcal{O}\left(\log(n)\right)$.
The overhead introduced in both schemes is only used for error estimation and it is non-trivial to use it for error correction.

We denote random variables with upper case letters and their realisations with the corresponding lower case letter, e.g., $A$ and $a$. For row-vectors we use bold faced letters, e.g., $\vec{X}$ or $\vec{x}$. For sets we use calligraphic symbols, e.g., $\set{X}=\{0,1\}$.
A probability parametrized by a deterministic variable is written as $\prob{A;\theta}$. The expectation with respect to the random variable $A$ is denoted by $\expect[A]{\cdot}$, and $\lfloor a \rfloor$ denotes the largest integer not greater than $a$. Finally, GF(2) denotes the Galois field of size $2$.

\section{System Model and Estimator}\label{sec:model}
Let $\mathcal{C}$ be the codebook of a binary linear code defined by a parity-check matrix $\pchk$ of dimension $m \times n$, i.e., $\mathcal{C}=\{\vec{x} \in \{0,1\}^n : \vec{x}\pchk^{\text{T}}=\vec{0}\}$. In particular, we focus on check-regular low-density parity-check (LDPC) codes \cite{GallagerLDPC} where every row of the parity-check matrix has constant weight $d$. We will use the term \emph{check node degree} to refer to the weight of a row. Let the row-vector $\vec{X}$ be a codeword of length $n$, i.e., $\vec{X} \in \mathcal{C}$. The elements $X_i$ ($i=1,\ldots,n$) of $\vec{X}$ are transmitted over a symmetric channel with binary input, i.e., the channel has input alphabet $\set{X}=\{0,1\}$ and an arbitrary output alphabet $\set{Y}$.

The receiver performs hard decisions on the received values $Y_i \in \set{Y}$ leading to $\hat{Y}_i \in \{0,1\}$ for all $i=1,\ldots,n$. Let $\rho$ denote the probability of error for these hard decisions, i.e., $\rho = \prob{\hat{Y}_i \neq X_i}$ and assume $0\leq \rho \leq \frac{1}{2}$. These hard decisions are used to compute the syndrome $\vec{S}$ by multiplying the row-vector of hard decisions $\hat{\vec{Y}}$ of the received vector with the parity-check matrix, i.e.,
\begin{align}
	\vec{S} = \hat{\vec{Y}} \pchk^{\text{T}},
\end{align}
where the operations are in GF(2).

In \cite{TotoZarasoa:2011p13281}, the authors claimed that the elements of $\vec{S}$ can be seen as the realisation of an (i.i.d.) Bernoulli process if the rows of $\pchk$ are linearly independent. 
Unfortunately, syndrome bits from check nodes with non-disjunct sets of variable nodes are (weakly) correlated, if $\rho\ne \frac{1}{2}$.
However, while not being precise, modelling the syndrome symbols as realisations of i.i.d. Bernoulli trials is a good approximation for a wide range of applications. An intuitive argument is that the realisations of the syndrome of a good channel code operating close to capacity should all be equally likely.\footnote{A similar argument can be made when binary block codes are used for lossless source coding: If the syndrome, i.e., the compressed data, would not be uniformly distributed then it could be compressed further.}

Motivated by this and by our numerical results, we model the elements of $\vec{S}$ as outcomes of a Bernoulli process with parameter $q$. Assuming that each row of the parity-check matrix has weight $d$ leads to
\begin{align}
q & = f_d(\rho) := \sum_{i \in \set{T}} {d\choose i}\rho^i (1-\rho)^{d-i}
		= \frac{1-(1-2\rho)^{d}}{2},\label{equ:p2q}
\end{align}
where $\set{T}$ is the set of positive odd integers not larger than $d$.
A proof for the last identity using the probability-generating function of the binomial distribution can be found in \cite[Lemma 1]{GallagerLDPC}.

Let $W=\weight(\vec{S})$ denote the Hamming weight of $\vec{S}$ which is a sufficient statistic for estimating $\rho$, and recall that $m$ is the length of $\vec{S}$. 
Using the i.i.d. approximation, the probability distribution of the Hamming weight of $\vec{S}$ is binomial
\begin{align}
	\prob{W=w} = {m\choose w}q^w(1-q)^{m-w},\label{equ:binomial}
\end{align}
and the maximum likelihood (ML) estimate for $\rho$ given a syndrome weight $w$ is
\begin{align}\label{equ:maxprob}
	\hat{\rho}(w) = \arg \max_{\rho'} \Bigg\{f_d(\rho')^w \left(1-f_d(\rho')\right)^{m-w}\Bigg\},
\end{align}
which can be solved analytically by setting the derivative (with respect to $\rho'$) of the argument to zero.
Equivalently, one can first use the ML estimator for $q$
\begin{align}
	\hat{q}(w) = \frac{w}{m},\label{equ:estq}
\end{align}
and use it with \eqref{equ:p2q} to obtain the estimate $\hat{\rho}$.
The final estimator in closed form is in both cases derived as
\begin{align}
	\hat{\rho}(w) = \left\{
	\begin{array}{lcl}
		\frac{1-\left(1-2\frac{w}{m}\right)^{\frac{1}{d}}}{2} &;& \frac{w}{m}\leq1/2\\
		\frac{1}{2} &;& \frac{w}{m}>1/2
	\end{array}
	\right..\label{equ:pest}
\end{align}

Note that $\hat{q}(w)$ in \eqref{equ:estq} is unbiased but due to the non-linearity of $f_d(\rho)$, cf. \eqref{equ:p2q}, the final estimator $\hat{\rho}(w)$ is biased.
 
For an irregular check node degree distribution it is straightforward to modify \eqref{equ:maxprob} to
\begin{align*}
	\hat{\rho}(\vec{w}) = \arg \max_{\rho'} \Bigg\{\prod_j f_{d_j}(\rho')^{w_j} \left(1-f_{d_j}(\rho')\right)^{m_j-w_j}\Bigg\},
\end{align*}
where $m_j$ is the number of check nodes of degree $d_j$ and $w_j$ is the weight of the part of the syndrome corresponding to check nodes of degree $d_j$. Unfortunately, there is no closed-form solution as in the regular case.

\section{Binary Symmetric Channel}\label{sec:bsc}
For the BSC, the hard decision $\hat{Y}_i$ is equivalent to the output of the channel $Y_i$ and the channel parameter of interest, the crossover probability of the BSC, is $\rho$. In this section we derive the bias and mean squared error (MSE) of this estimator using the i.i.d. approximation for the syndrome bits and compare the MSE to the Cram\'er-Rao bound.

\subsection{Bias and Mean Squared Error}
The estimator of $\rho$ is biased and its mean is
\begin{align}
\mu(d, \rho, m)&=\expect[W]{\hat{\rho}(W)}\label{equ:bscmean}\\
&\hspace{-1.5cm} = \frac{1}{2} - \frac{1}{2}\sum_{w=0}^{\lfloor m/2 \rfloor} {m \choose w}f_d(\rho)^w(1-f_d(\rho))^{m-w}\left( 1-2\frac{w}{m}\right)^{\frac{1}{d}},\nonumber
\end{align}
with $f_d(\rho)$ defined in \eqref{equ:p2q}.
The bias of the estimator is therefore
\begin{align}
	B(d,\rho,m) &= \mu(d,\rho,m)-\rho.\label{equ:bscbias}
\end{align}

The mean squared error of the estimator $\hat{\rho}$ is
\begin{align}\label{equ:bscmse}
\text{MSE}(d,\rho,m) =& \expect[W]{\left(\hat{\rho}(W)-\rho\right)^2}\\
=& \frac{1}{4}-2\rho\mu(d,\rho,m)+\rho^2\nonumber\\
&\hspace{-1.5cm}+\frac{1}{4}\sum_{w=0}^{\lfloor m/2 \rfloor}{m \choose w}f_d(\rho)^w(1-f_d(\rho))^{m-w}\Bigg(\left(1-2\frac{w}{m}\right)^{\frac{2}{d}}\nonumber\\
&-2\left(1-2\frac{w}{m}\right)^{\frac{1}{d}}\Bigg).\nonumber
\end{align}

When evaluating \eqref{equ:bscmean}, \eqref{equ:bscbias} and \eqref{equ:bscmse} for large $m$ it is useful to approximate the binomial distribution \eqref{equ:binomial} with a Poisson and Gaussian distribution for small and large values of $\rho$, respectively.

\subsection{Cram\'er-Rao Bound}
For a finite number of samples, the estimator \eqref{equ:pest} is biased and therefore, its variance is lower bounded by the \emph{biased} Cram\'er-Rao bound \cite[Chapter 3.6]{kay1998fundamentals}:
\begin{align}
	V(d,\rho,m) &\geq \frac{\left( 1+\frac{\partial}{\partial \rho}B(d,\rho,m) \right)^2}{\mathcal{I}(\rho)}\\
	&=\frac{\left(\frac{\partial}{\partial \rho}\mu(d,\rho,m) \right)^2}{\mathcal{I}(\rho)}\nonumber,
\end{align}
and hence the mean squared error is bounded as
\begin{align}\label{equ:bscmsebound}
	\text{MSE}(d,\rho,m) \geq \frac{\left(\frac{\partial}{\partial \rho}\mu(d,\rho,m) \right)^2}{\mathcal{I}(\rho)} + B^2(d,\rho,m).
\end{align}

The Fisher information $\mathcal{I}$  the syndrome $\vec{S}$ carries about the BSC parameter $\rho$ is
\begin{align}\label{equ:fisher}
\mathcal{I}(\rho) =&
-\expect[\vec{S}]{\frac{\partial^2}{\partial \rho^2}\log \prob{\vec{S};\rho}}\\
=&-\expect[W]{\frac{\partial^2}{\partial \rho^2}\log \left(q^{W}(1-q)^{m-W}\right)}\nonumber\\
=& \frac{4md^2(1-2\rho)^{2d-2}}{1-(1-2\rho)^{2d}}.\nonumber
\end{align}

The estimator \eqref{equ:pest} is a ML estimator; so it is asymptotically efficient: for a large number of samples its bias converges to zero and its MSE converges to the unbiased Cram\'er-Rao bound which is the inverse of the Fisher information \eqref{equ:fisher}.

\subsection{Analysis}\label{sec:bscnumerical}
\begin{figure}[ht]
  \begin{center}
    \includegraphics[width=\figwidth]{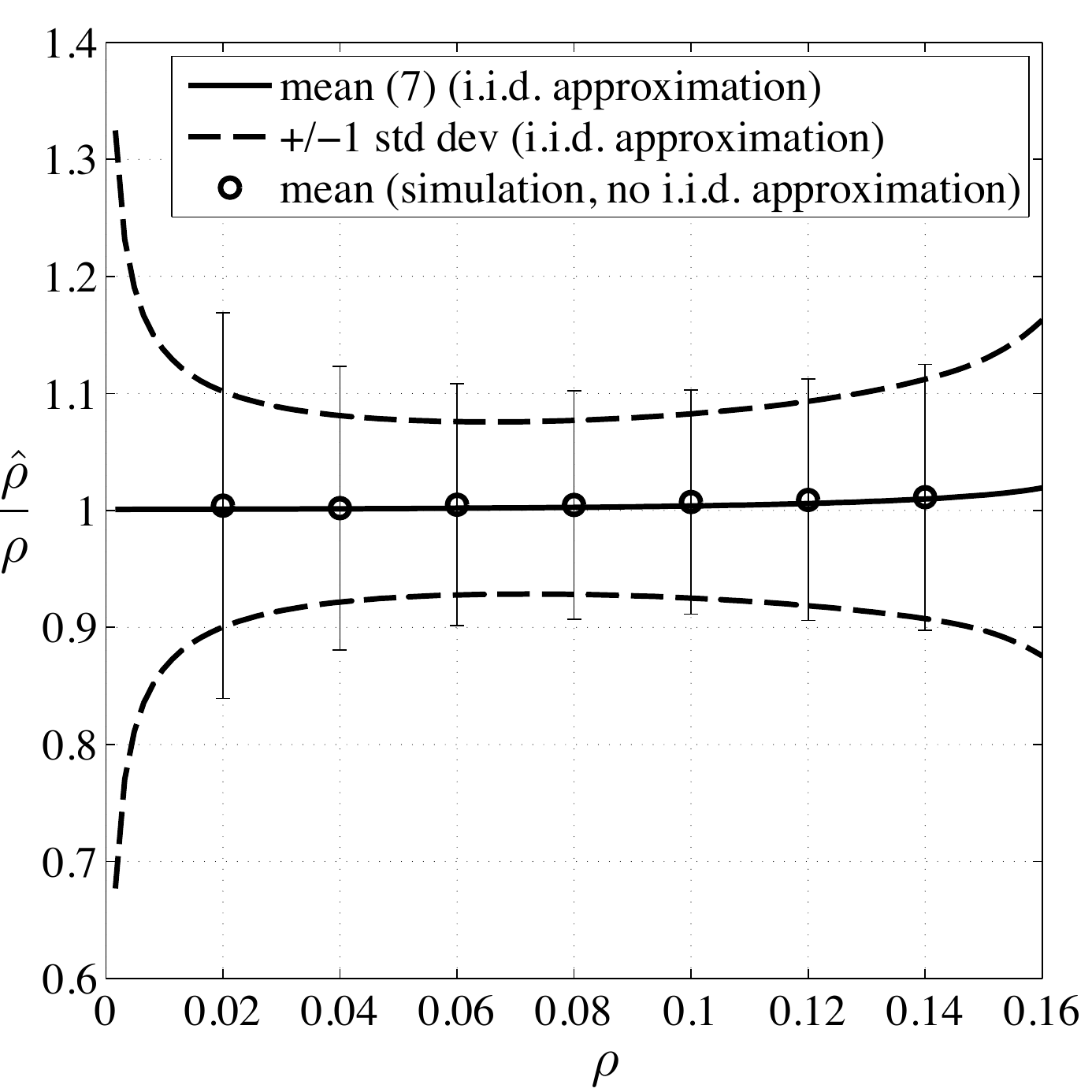} 
    \caption{Normalised mean and standard deviation of the BSC estimator \eqref{equ:pest} for $d=6$ and $m=1,000$. Markers denote simulation results where the error bar corresponds to one standard deviation.}
    \label{fig:rhomeanrel}
  \end{center}
\end{figure}

\begin{figure}[ht]
  \begin{center}
    \includegraphics[width=\figwidth]{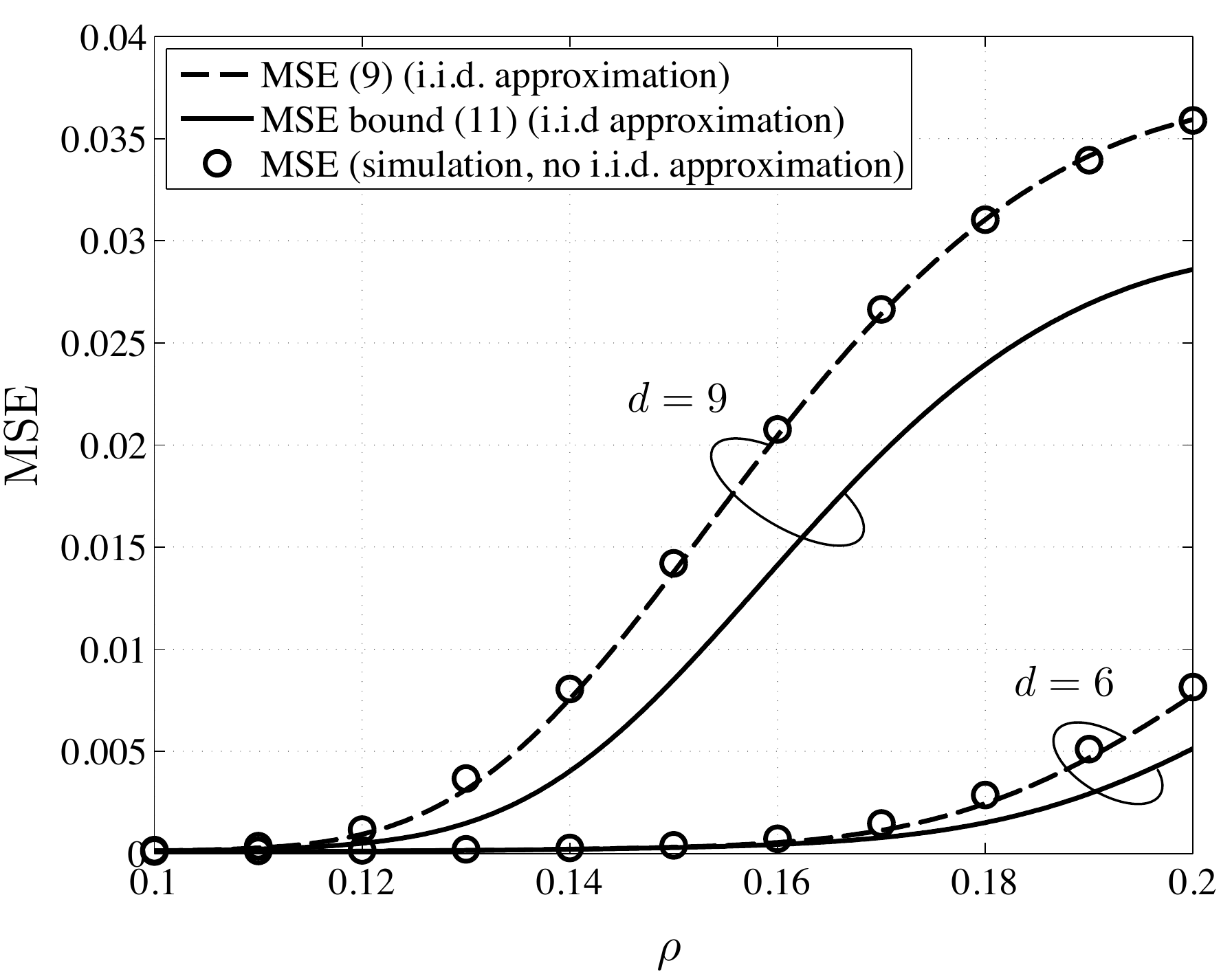} 
    \caption{MSE \eqref{equ:bscmse} of the BSC estimator and its lower bound \eqref{equ:bscmsebound} for check nodes of degree $d=6$ and $d=9$ for $m=1,000$. Simulation results are for regular LDPC codes with variable nodes of degree $3$.}
    \label{fig:rhomse}
  \end{center}
\end{figure}

In Fig.~\ref{fig:rhomeanrel} we show the normalised estimator mean and standard deviation as a function of the parameter $\rho$ for a check node degree $d=6$ and $m=1,000$ check nodes. The analytical mean of the estimator is close to the true parameter $\rho$ and the normalised standard deviation increases for small and large $\rho$. 
Note, that the MSE \eqref{equ:bscmse} approaches zero as $\rho$ tends to zero (see also Fig.~\ref{fig:rhomse}); so the divergence of the normalized standard deviation at $\rho = 0$ is purely due to the division by $\rho$.

In addition to the analytical results, which rely on the independence assumption, simulation results are shown as markers and error bars. For the simulation we used a regular\footnote{A simulation of an LDPC code with $d = 6$ and an irregular \emph{variable} node degree distribution showed only negligible differences.} LDPC code with variable nodes of degree $3$ and check node degree $d=6$. While the simulated mean is close to the analytical result, the simulated normalised standard deviation is larger than the analytical result due to the violation of the independence assumption of the syndrome symbols.

In Fig.~\ref{fig:rhomse} we compare the MSE of the estimator \eqref{equ:bscmse} with its lower bound \eqref{equ:bscmsebound}. We assume again $m=1,000$ and check node degrees $d=6$ and $d=9$. Due to the relatively small number of check nodes there is a relatively large gap to the bound. In addition, we observe that a higher check node degree leads to a significant increase of the MSE.

\begin{figure}[ht]
  \begin{center}
    \includegraphics[width=\figwidth]{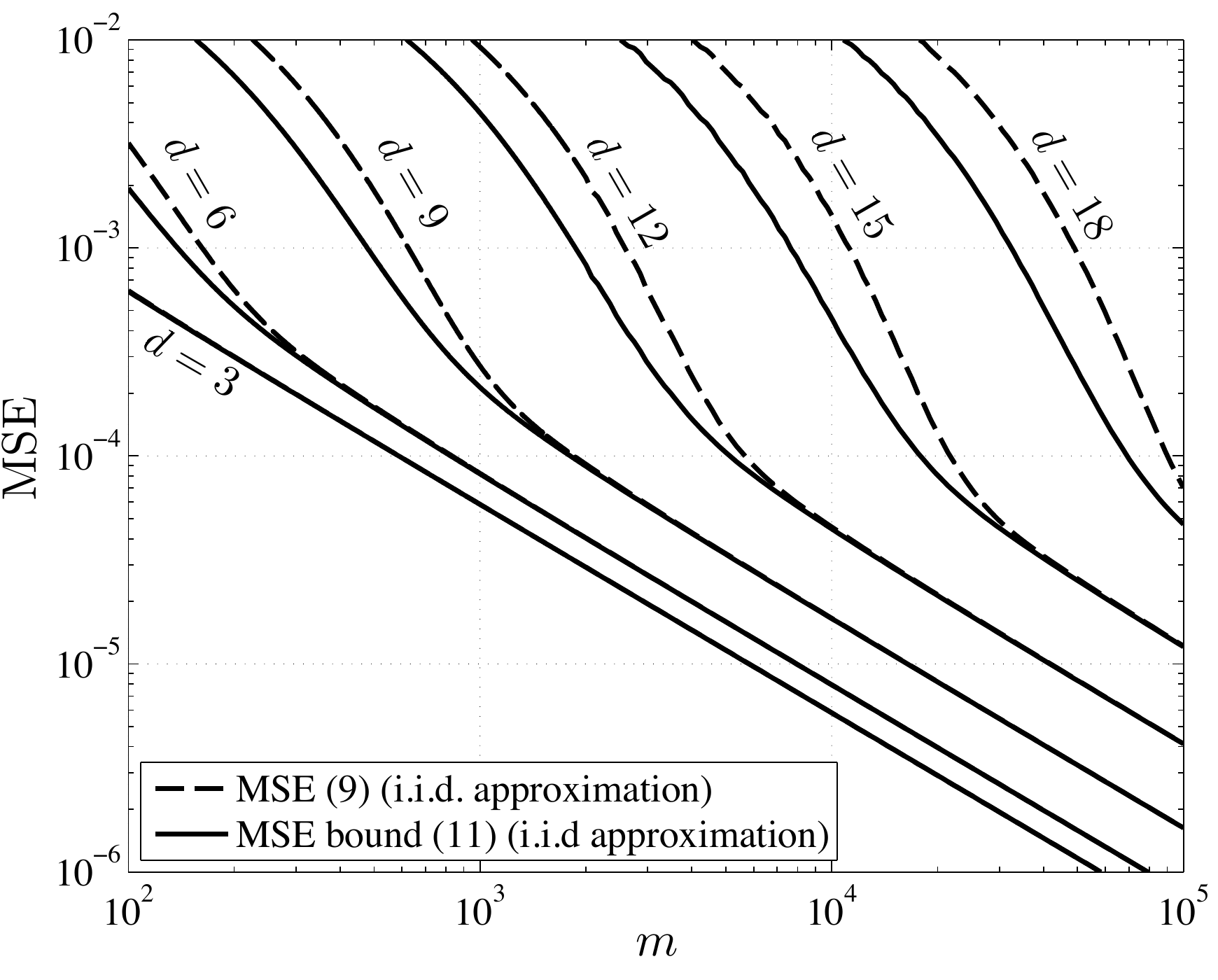} 
    \caption{MSE \eqref{equ:bscmse} and its lower bound \eqref{equ:bscmsebound} as a function of the number of check nodes $m$ and their degree $d$ for $\rho=0.11$.}
    \label{fig:rhodm}
  \end{center}
\end{figure}

Finally, we show the MSE as a function of the number of check nodes $m$ and the check degree $d$ for $\rho=0.11$ in Fig.~\ref{fig:rhodm}.
For small check node degrees a relatively small number of check nodes leads to a small MSE. For a large number of check nodes, the curves in Fig.~\ref{fig:rhodm} approach the inverse of the Fisher information, i.e., the unbiased Cram\'er-Rao bound. 
This is important in practice as one can use \eqref{equ:fisher} to obtain an upper bound on the check node degree that can be used for code optimisation if the code is to be used for channel estimation.

The number of check nodes $m$ directly enters our analytical results, i.e., for large $m$ the MSE tends to zero as the inverse of $m$.  
The number of variable nodes $n$ does not enter our results, however, it influences the accuracy of the independence assumption: for a fixed number of check nodes $m$ and fixed check node degree $d$ increasing the codeword length $n$ reduces the average variable node degree and thus the correlation between parity bits. Consequently, simulation result are closer to the analytical results for the i.i.d. case.

\section{Gaussian Channel with Binary Input}\label{sec:awgn}
Consider now a BI-AWGN channel with transmit energy $E_s$ and noise variance $\sigma^2$. The signal-to-noise ratio (SNR) in logarithmic scale is defined as
\begin{align}
	\gamma = 10\log_{10}\frac{E_s}{2\sigma^2}.
\end{align}
Estimating the SNR directly from the received signal is an established problem (e.g.,\cite{Pauluzzi2000,Alagha2001}). We are interested in estimating the SNR via the syndrome, i.e., the receiver performs a hard decision of the received signal which converts the overall channel to a BSC. From the crossover probability of the BSC we derive the SNR of the BI-AWGN channel.

The probability of error of a hard decision is $\rho = Q(\gamma)$, where
\begin{align}\label{equ:qfunc}
Q(\gamma) = \frac{1}{\sqrt{2\pi}}\int_{10^{\gamma/10}}^{\infty} e^{-\frac{\psi^2}{2}} d\psi.
\end{align}

Therefore, if we again approximate the syndrome bits to be i.i.d., the ML estimator for $\gamma$ is the inverse of \eqref{equ:qfunc}
\begin{align}
	\hat{\gamma}(w) &= 10\log_{10}\left(Q^{-1}\left(\hat{\rho}(w)\right)\right).
\end{align}
This estimator diverges for arguments $w=0$ and $w\geq m/2$. Therefore, we consider a modified estimator that restricts the possible values for the estimate to the interval $[\gamma_{\text{min}}, \gamma_{\text{max}}]$.
\begin{align}
	\tilde{\gamma}(w) = \left\{
	\begin{array}{lcl}
		\gamma_{\text{min}} &;& \hat{\gamma}(w)<\gamma_{\text{min}}\\
		\hat{\gamma}(w) &;& \gamma_{\text{min}}\leq\hat{\gamma}(w)\leq\gamma_{\text{max}}\\
		\gamma_{\text{max}} &;& \hat{\gamma}(w)>\gamma_{\text{max}}
	\end{array}
	\right..\label{equ:snrest}
\end{align}

For the following analysis, we numerically calculate mean, bias, and MSE of this estimator \eqref{equ:snrest}, i.e. 
\begin{align}
\mu(d, \gamma, m)&=\expect[W]{\tilde{\gamma}(W)},\\
  B(d,\gamma,m) &= \mu(d,\gamma,m)-\gamma, \\
\text{MSE}(d,\gamma,m) &= \expect[W]{\left(\tilde{\gamma}(W)-\gamma\right)^2}.
\label{equ:snrmse}
\end{align}


\subsection{Analysis}
\begin{figure}[ht]
  \begin{center}
    \includegraphics[width=\figwidth]{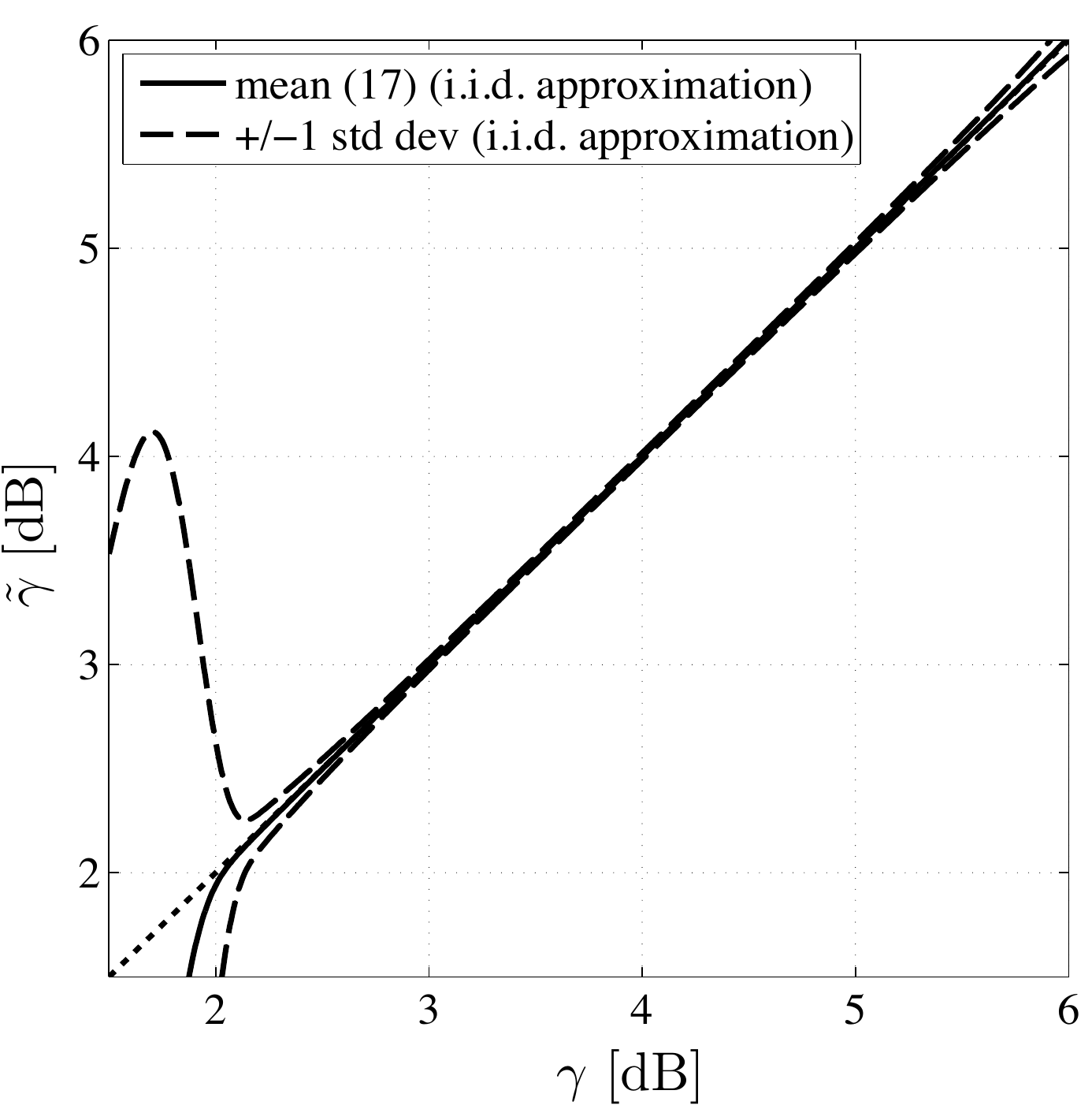} 
    \caption{Estimator mean and standard deviation for $m=10,000$ and check nodes of degree $d=30$ ($\gamma_{\text{min}}=-10$~dB, $\gamma_{\text{max}}=10$~dB).} 
    \label{fig:snrmean}
  \end{center}
\end{figure}

We are interested in the range of SNR for which the syndrome based estimator delivers accurate estimates. For this purpose we show the mean of the estimator output in Fig.~\ref{fig:snrmean} for $m=10,000$ and $d=30$ (such a check node degree would be typical for a regular LDPC code of rate $0.9$ with variable node degree $d_v=3$). The SNR interval was restricted to $\gamma_{\text{min}}=-10$~dB and $\gamma_{\text{max}}=10$~dB. In addition, the confidence interval for one standard deviation is shown. 
We see that for the given parameters the estimator works well above $2$~dB. This is below the capacity-threshold of $2.74$~dB for a code with rate 0.9 and hence the estimator provides accurate results in a range where the error correcting code will be used.

We note that the accuracy of the estimator degrades for low SNR. This is due to the fact that our estimator operates on hard decisions which is suboptimal, especially at low SNR.
In general, estimating the SNR directly from the channel outputs \cite{Pauluzzi2000} will lead to more accurate results when compared to our estimator. However, our estimator is solely based on hard-decisions which has a complexity advantage when compared to soft-decision based estimators.

\begin{figure}[ht]
  \begin{center}
    \includegraphics[width=\figwidth]{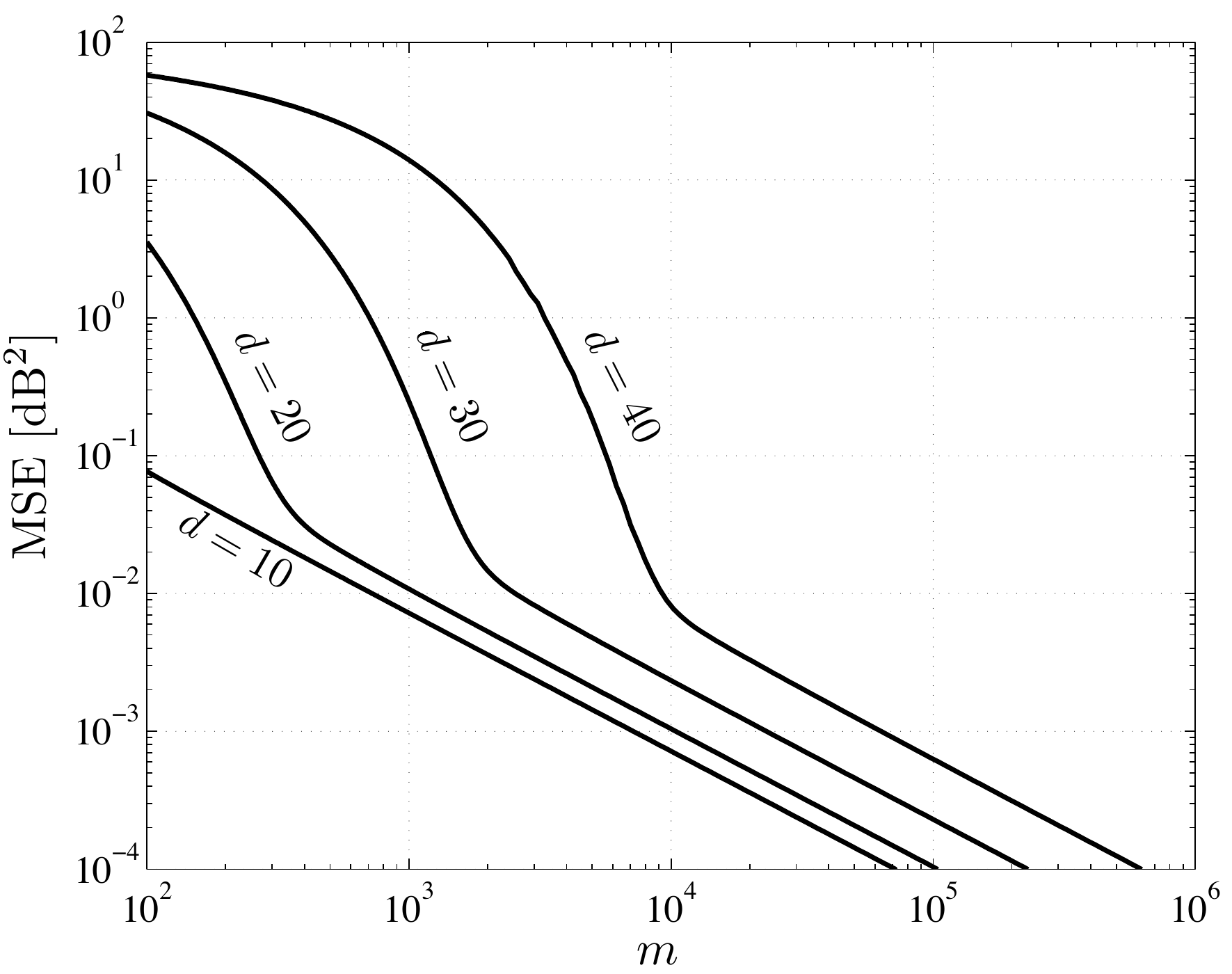} 
    \caption{MSE \eqref{equ:snrmse} as a function of the number of check nodes $m$ and their degree $d$ for $E_s/2\sigma^2=2.5$ dB ($\gamma_{\text{min}}=-10$~dB, $\gamma_{\text{max}}=10$~dB).}
    \label{fig:snrdm}
  \end{center}
\end{figure}

Similar to the BSC case, we investigate the influence of the number of check nodes $m$ and the check degree $d$ on the MSE. The results are shown in Fig.~\ref{fig:snrdm} for $E_s/2\sigma^2 = 2.5$ dB where we observe a similar behaviour as for the BSC.

\section{Conclusions}
We derived analytical expressions for the properties of a maximum likelihood estimator for the crossover probability of a BSC based on the syndrome bits (assumed to be i.i.d.) of a linear code. The accuracy of the estimator is determined by the number of parity-checks and by their degree.
This result is important for code design as it puts an upper bound on the check node degree that can be used for code optimisation.

Finally, we used the BSC based estimator to estimate the SNR of a BI-AWGN channel. This estimator is limited to a specific range of SNR as outside this range, its mean deviates from the channel parameter and its MSE increases significantly. Similar to the BSC case we analysed the influence of the number of check nodes and their degree.

\section*{Acknowledgements}
The authors would like to thank Priska Lang for fruitful discussions and the anonymous reviewers for comments that helped to improve the paper.
\bibliographystyle{IEEEtran}
\bibliography{references}

\end{document}